# Unusual Electrical Conductivity Driven by Localized Stoichiometry Modification at Vertical Epitaxial Interfaces

Wenrui Zhang,[1,2,*] Shaobo Cheng,[3] Christopher M Rouleau,[2] Kyle P. Kelley,[2] Jong Keum,[2,4] Eli Stavitski,[5] Yimei Zhu,[3] Matthew F. Chisholm,[2] Zheng Gai,[2] and Gyula Eres[1,2,*]

[1]Materials Science and Technology Division, Oak Ridge National Laboratory, Oak Ridge, Tennessee 37831, United States
[2]Center for Nanophase Materials Sciences, Oak Ridge National Laboratory, Oak Ridge, Tennessee 37831, United States
[3]Condensed Matter Physics and Materials Science, Brookhaven National Laboratory, Upton, New York 11973, United States
[4]Neutron Scattering Division, Oak Ridge National Laboratory, Oak Ridge, TN 37831, United States
[5]National Synchrotron Light Source II, Brookhaven National Laboratory, Upton, New York 11973, United States

**Abstract**

Precise control of lattice mismatch accommodation and cation interdiffusion across the interface is critical to modulate correlated functionalities in epitaxial heterostructures, particularly when the interface composition is positioned near a compositional phase transition boundary. Here we select $La_{1-x}Sr_xMnO_3$ (LSMO) as a prototypical phase transition material and establish vertical epitaxial interfaces with NiO to explore the strong interplay between strain accommodation, stoichiometry modification, and localized electron transport across the interface. It is found that localized stoichiometry modification overcomes the plaguing dead layer problem in LSMO and leads to strongly directional conductivity, as manifested by more than three orders of magnitude difference between out-of-plane to in-plane conductivity. Comprehensive structural characterization and transport measurements reveal that this emerging behavior is related to a compositional change produced by directional cation diffusion that pushes the LSMO phase transition from insulating into metallic within an ultrathin interface region. This study explores the nature of unusual electric conductivity at vertical epitaxial interfaces and establishes an effective route for engineering nanoscale electron transport for oxide electronics.

## 1. Introduction

The electronic structure of the interface between two correlated oxides is subject to a variety of influences that can create intriguing electronic phases and functionalities, including charged domain walls in ferroelectrics,[1] high mobility two-dimensional electron gas (2DEG) systems,[2,3] and large magnetoelectric coupling in multiferroics.[4,5] To engineer interface-driven functionalities, a key prerequisite is precise control of the chemical profile and the local structural ordering at the interface. The lattice mismatch at epitaxial interfaces is a critical design parameter that is fundamentally connected to atomic arrangements and lattice distortions across the interface.[6,7] While a large lattice mismatch is usually accommodated by the formation of misfit dislocations illustrated in **Scheme 1a**,[8,9] it can have an alternative effect by modifying the local composition near the interface region illustrated in Scheme 1b.[10,11] Such diffusion-driven rearrangements can have particularly dramatic impact on physical properties when the chemical fluctuation occurs at composition-controlled insulator-metal phase boundaries. For example, in a model correlated oxide $La_{1-x}Sr_xMnO_3$ (LSMO), a subtle change of the Sr doping level ($x$) from 0.5 to 0.44 results in an abrupt phase transition from an antiferromagnetic (AFM) insulator to a ferromagnetic (FM) metal below 250 K (Scheme 1c).[12,13] On the other hand, the interface coupling can extend and penetrate deep into both sides of the interface, creating new phenomena known as critical thickness or dead layer effects in LSMO, which adversely affects the metallic and ferromagnetic functionalities of the LSMO thin films.[14-17] Therefore, it is critical to develop effective ways to control the interface transformations towards the desired direction to improve the functional response.

In conventional interfaces with a parallel film-substrate geometry, the buried interface nature makes it challenging to decouple of the interface transport from the bulk. Alternatively, a highly attractive device architecture is generated in epitaxial vertically aligned nanocomposite (VAN) thin films formed by self-assembled phase segregation where the interfaces are flipped perpendicular to the film surface.[21-23] The vertical epitaxial interfaces in such VAN structures enables more effective strain coupling and demonstrates profound impact on magnetic coupling,[24,25] magnetoresistance[26] and electrical and ionic transport.[27] Specifically, we previously reported on establishment of a well-ordered VAN architecture using NiO and ferromagnetic LSMO ($x = 0.3$), and realized perpendicular exchange bias for tailoring magnetotransport.[25]

Taking advantage of the ultrahigh interface density and the extremely large lattice mismatch of the LSMO-NiO VAN system, this work systematically varies the Sr doping level $x$ in LSMO across its insulator-to-metal compositional phase boundary ($x = \sim0.47$), and explores the correlation between strain accommodation, stoichiometry modification, and interfacial electron transport. We observe a dramatic enhancement of electrical conductivity in the out-of-plane (OP) direction in LSMO-NiO VAN films. This conductivity is unusual for two reasons. First, it occurs in a composition range $x \geq 0.5$ where LSMO thin films are nominally insulating. Second, the existence of enhanced interface conductivity of LSMO is remarkable because LSMO thin films are known for the opposite effect, the degradation of interface conductivity caused by dead-layer formation.[15-17] We refer to this unexpected conductivity enhancement as an "inverse dead-layer" effect because it occurs at interfaces where the constituting LSMO is nominally in the insulating phase. In-depth structural and electrical characterizations strongly suggest that the dramatic directional conductivity enhancement in the LSMO-NiO VAN films arises from an insulator-to-metal phase transition that is driven by spontaneous La enrichment at the vertically aligned interface region.

## 2. Results and Discussion

2.1 Epitaxial vertical interfaces and strain evolution

The LSMO-NiO nanocomposite thin films or VAN structures were fabricated by pulsed laser deposition (PLD) on single crystal SrTiO$_3$ (STO) (001) substrates. The VAN structures form by spontaneous phase separation between LSMO and NiO during PLD film growth from composite targets consisting of a fixed 3:2 molar ratio of LSMO and NiO, while *x* in LSMO is systematically tuned. The variation of *x* has negligible effect on the VAN film microstructure, which is examined by scanning transmission electron microscopy (STEM) in the high-angle annular dark-field (HAADF) mode. As seen from the plan-view STEM image in **Figure 1a**, the VAN films consist of LSMO as the majority phase, and NiO as the minority phase with mixed equiaxed and elliptical morphologies viewed from their long axis. This gives an ultrahigh pillar density in the order of $10^{15}$ m$^{-2}$ using an estimate of ~5 nm for the average diameter and of 15 nm for the pitch between NiO nanopillars. These NiO nanopillars form continuous vertical interfaces along their entire length through the LSMO film shown in Figure 1b. The atomic-resolution STEM images and its crystallographic model interpretation in Figure 1c reveal the growth of a vertical epitaxial interface of high perfection between LSMO and NiO.

X-ray diffraction (XRD) results confirm the epitaxial growth of LSMO-NiO VAN films and determine the residual stain states. The VAN films are oriented along the (00*l*) axis as its out-of-plane (OP) direction (Figure S1). The biaxial strain imposed by the lattice mismatch with the substrate is largely relaxed with increasing films thickness. The residual OP strain ($\varepsilon_{OP}$) for 40-nm-thick single-phase LSMO and NiO films is less than 0.5%, which is calculated using c-axis lattice parameters of LSMO ($c_{LSMO}$) and NiO ($c_{NiO}$) versus their bulk values of 0.417 nm for NiO, and 0.387 nm for LSMO. On the other hand, the vertical LSMO-NiO interface effectively redistributes a fraction of the mismatch strain to the constituent phases, leading to apparently compressive strain for $c_{NiO}$ and smaller tensile strain for $c_{LSMO}$ from the XRD analysis of (002) peak shift (Figure 1d). As $x$ changes from 0.3 to 0.5 across the phase transition boundary, $c_{LSMO}$ varies from 0.3879 nm to 0.386 nm, while $c_{NiO}$ changes more drastically from 0.4088 nm to 0.4029 nm. Reciprocal space maps (RSMs) in Figure S2 near the asymmetric (103) diffraction peak confirm cube-on-cube growth of both single-phase LSMO, and NiO, as well as the VAN films on STO (001) substrates. The RSMs show that vertical interface coupling predominantly affects the film c-axis lattice parameter, causing the LSMO (103) peak to shift along the out-of-plane direction, as illustrated for LSMO-NiO films with $x = 0.3$ and $x = 0.5$ in Figure 1e and 1f. Note that the bulk lattice parameter of LSMO is very close in the composition range of interest ($0.3 \leq x \leq 0.6$),[28] thus the nominal strain from the lattice mismatch between LSMO and NiO is expected to be independent of $x$ across this composition range. However, following epitaxial growth of the LSMO-NiO VAN films, the residual strain states of LSMO and NiO were found to evolve differently as a function of $x$, suggesting a strong interplay between the chemical composition and vertical interface coupling.

2.2 Localized interface structure

To probe the structural and chemical profile of the LSMO-NiO interface, we perform atomic-resolution STEM imaging combined with electron energy loss spectroscopy (EELS). The plan-view STEM image in **Figure 2a** shows that the LSMO matrix matches NiO nanopillars with two dominant relations, $[100]_{LSMO}\|[100]_{NiO}$ and $[110]_{LSMO}\|[100]_{NiO}$, thus forming two major facets ({100} and {110}) on NiO nanopillars. Geometric phase analysis (GPA) in Figure 2b and 2c shows the LSMO and NiO interior lattices present uniform strain contrast, while their interface displays distinct strain concentrations as a result of dislocation formation. Analysis of the lattice parameter variation near the LSMO-NiO interface in Figure 2d reveals a subtle strain accumulation and relaxation process at the interface region. The lattice parameter for LSMO and NiO initially show opposite distortion, compressive for LSMO and tensile for NiO, before relaxing to steady state values.

The chemical environment of the interface region is clearly different from the bulk region of the constituent phases. Despite the overlapping La-$M_4$ and Ni-$L_3$ edges, the phase segregation between LSMO and NiO is clearly visible from the distinct variation of La-$M_5$ and Ni-$L_2$ edges (Figure 2e). The oxidation state of La and Ni remain unchanged at 3+ and 2+. By contrast, the Mn-$L_3$ edge energy that is constant far from the interface starts to reduce gradually as the probe area approaches the interface shown in Figure 2f. This energy shift indicates a reduction in the Mn oxidation state, which is related to La segregation at the interface region.[29,30] The intensity variations in the LSMO surrounding the NiO nanopillars seen in Figure 2a are consistent with local change in La content. As a result, a localized region with modified stoichiometry arises at the interface that extends 2-3 nm into LSMO. This phenomenon is persistent for LSMO-NiO VAN films and is confirmed for film compositions near and above the phase transition boundary at $x = 0.47$ and $x = 0.5$, respectively. Since the stoichiometry change occurs only in the local interface region, it has negligible effect on the overall oxidation state of Mn. The X-ray absorption near-edge structure (XANES) results in Figure S3 show nearly identical absorption features for LSMO-NiO VAN films with $x = 0.44$ and $x = 0.5$ confirming that the oxidation states remain unchanged in the bulk of the films.

2.3 Magnetic phase transition and directional electrical transport

The ultrahigh density of self-assembled vertical interfaces has a profound influence on the magnetic and electronic states of the LSMO-NiO VAN films. The bulk phase diagram shows that a magnetic phase transition can be induced by modifying the composition of LSMO near the phase boundary region at $x = \sim 0.47$.[12,13] Changing $x$ from 0.3 to 0.5, reduces the saturation magnetization ($M_s$) from 255 to 78 emu/cc, and increases the coercive field ($H_c$) from 274 to 642 Oe measured at 10 K (Figure S4). A comparison between single-phase LSMO and LSMO-NiO VAN films reveals that incorporation of NiO nanopillars leads to frustrated magnetization and enhanced coercive field ($H_c$), suggesting effective FM-AFM coupling at the vertical interface.[31,32]

The temperature dependent resistivity (*R-T*) of LSMO-NiO VAN films in Figure 3 departs from that of the single-phase LSMO film and depends strongly on $x$. We first discuss the *R-T* results measured along the IP direction shown in **Figure 3a**. The single-phase LSMO film for $x = 0.3$ exhibit metallic behavior for the entire measurement temperature from 10 to 350 K. Incorporation of NiO nanopillars introduces spin scattering centers [33] and lowers the metal-to-insulator transition temperature for LSMO with $x$ in the range of 0.3-0.47. As $x$ increases above the FM-AFM phase transition boundary (~0.47), the LSMO-NiO VAN films exhibit only insulating behavior, with the film resistance increasing above the instrument measurement limit at low temperatures. These data indicate that in the IP direction the LSMO matrix dominates electrical transport.

A more profound result is a dramatic enhancement of electrical conductivity in the OP direction in the LSMO-VAN films that consist of nominally insulating LSMO and NiO. The single-phase LSMO film with $x = 0.3$ is used as calibration points for establishing the baseline *R-T* relationships, which are almost identical between IP and OP. Such negligible conductivity anisotropy is also seen in LSMO-NiO VAN films with $x$ ranging between 0.3 and 0.47. This can be understood that the LSMO in this composition range is metallic, so that it dominates the conductivity channels in both directions, as illustrated in Figure 3d. As $x$ crosses the composition boundary into the insulating phase region, a dramatic change in the conductivity anisotropy of the LSMO-NiO VAN films appears. The summary plot in Figure 3c compares the IP and OP conductivity as a function of composition across the FM-metal-AFM-insulator phase boundary. The LSMO-NiO VAN films exhibit more than three orders of magnitude enhancement in the OP conductivity for $x \geq 0.47$, where according to the bulk phase diagram LSMO is nominally an AFM insulator. The dramatic enhancement of the OP conductivity is attributed to the formation of vertical conduction channels schematically illustrated in Figure 3e. Since NiO and LSMO for $x \geq 0.5$ are both insulators in the pure films, it must be assumed that the new conductivity component originates at the vertical interfaces between LSMO and NiO shown by the HAADF-STEM in Figure 1b and 1c.

2.4 Nanoscale mapping of localized electron transport

We used scanning tunneling microscopy/spectroscopy (STM/S) to identify and characterize the nature of the conductivity regions in the LSMO-NiO VANs. STM provides atomic resolution imaging to locate and probe the spatial distribution of localized conductive states together with STS at the LSMO-NiO interface. We selected two representative LSMO-NiO VAN films with $x = 0.44$ for metallic and $x = 0.50$ for insulating phase for performing STM/S characterization at 77 K. The STM image in **Figure 4a** shows the identical surface morphology corresponding to the plan-view STEM image in Figure 1a. The film surface is similar to ordinary heteroepitaxial oxide films characterized by steps and smooth terraces,[34] except that the terraces are uniformly decorated with small holes in the LSMO surface, about 0.5 nm deep, shown in the line profile in Figure 4b. The hole regions in the STM images are identified as the NiO pillars based on the similarity of the density, the size, and the spacing of the features observed by STEM imaging in

Fig 1a. Figure 4c presents a two-dimensional intensity map of spatially resolved differential conductance (dI/dV) as a function of the bias voltage for a representative region with a NiO nanopillar embedded in the LSMO matrix for $x = 0.5$. Representative dI/dV spectra of the LSMO, the interface, and the NiO are overlaid in the same contour plot. As the STM tip scans from the LSMO matrix toward the NiO pillar region the gap edge reaches minimum indicating that the highest density of states is at the interface.[35-36] This signal is representative of localized conductivity at the interface and it is observed repeatedly as the STM tip crosses out of the NiO nanopillar region. In contrast, for VANs with $x = 0.44$ where LSMO is in the metallic composition region the magnitude of the gap edge in Figure 4d increases continuously from LSMO to NiO characteristic of electronic transitions from a metallic to an insulating state, indicating that no conducting states are present at the interface.

Complementary macroscopic electrical transport and nanoscale STM/S measurements reveal the crucial role of vertical epitaxial interfaces in supporting robust OP conductivity in the LSMO-NiO VAN films. The formation of vertical interface conductivity contrasts to the conventional dead layer problem that persistently plagues the performance of LSMO thin films, which can be referred as an inverse dead layer effect. Based on STEM, EELS and STM/S results, the vertical interface width is in the order of 3 nm and is very close to the dead-layer thickness in LSMO films. The origin of strong vertical conductivity is understood by the insulator-to-metal transition induced by spontaneous evolution of the insulating LSMO near the interface region. The STEM/EELS characterization identifies consistent reduction of the Mn oxidation state in LSMO as approaching to the interface region, which translates to preferential La accumulation and effective reduction of Sr content in LSMO at the interface.[29,30] The local modification of the Sr doping level can be enough to induce an insulator-to-metal phase transition at the interface if the LSMO composition is sufficiently close to the phase transition boundary. This is seen for LSMO-NiO VAN films with $x$=0.5, where the interface stoichiometry modification pushes LSMO across the phase boundary into the metallic composition range.

Although cation diffusion is well known for blurring interface sharpness and adversely affecting near-interface properties,[10,37] observation of such phenomena that produce desirable effects at diffusely broadened epitaxial interfaces are rather rare. This study emphasizes the positive role of interface stoichiometry modification in establishing robust conductive channels at the vertical interfaces formed between two nominally insulating oxides. This is rather surprising for LSMO, which is more known for the prevalence of dead layer formation that typically compromises interface conductivity. Understanding the detailed mechanisms driving composition evolution at vertical interfaces requires further theoretical and experimental effort, the results of which could ultimately yield novel mechanisms for controlling interface structure related to correlated functionalities. We note that the large lattice mismatch between LSMO and NiO can play a critical role in such mechanism, since lattice strain is known to have a strong effect on cation migration.[38,39] Finally, we highlight the key practical benefits of epitaxial vertical interfaces resulting from this study that include self-assembly, ultrahigh interface density, effective strain coupling and a device-friendly geometry for vertical transport.

## 3. Conclusion

We have demonstrated unexpected electrical conductivity supported by vertical epitaxial interfaces in self-assembled LSMO-NiO VAN films with various Sr doping levels. Temperature-dependent electric transport measurements reveal robust OP conductivity that is more than three orders of magnitude larger than in-plane ones in nanocomposite films consisting of nominally insulating LSMO and NiO. Atomic resolution STEM/EELS characterization identify slight stoichiometry modification at vertical epitaxial interfaces, which suggests a localized insulator-to-metal transition in LSMO for creating vertically conductive channels, instead of causing the plaguing dead layer problem in LSMO. STM/S measurements explore the spatial distribution of conductive states with nanoscale lateral resolution and identify the vertical interface as the origin of the robust OP conductivity. This study reveals unusual local electrical transport that qualifies as an inverse dead-layer effect at LSMO-NiO epitaxial vertical interfaces and provides insights into modulating physical properties of correlated oxides using vertically oriented epitaxial interfaces.

## 4. Experimental Section

**Thin film growth.** The LSMO-NiO VAN and the single-phase LSMO and NiO films were grown on STO(001) single-crystal substrates by PLD using a KrF excimer laser ($\lambda$= 248 nm) with fluence of 2 J/cm$^2$ and a repetition rate of 5 Hz. All films were grown at 200 mTorr $O_2$ and 700 °C. The films were post-annealed in 200 Torr of $O_2$ at 700 °C to ensure full oxidation, and cooled down to room temperature at a cooling rate of 20 °C/min. For out-of-plane transport measurements, the films were grown on 0.5% Nb-doped STO(001) single-crystal substrates. The film composition was varied by using composite laser ablation targets with different composition.

**Structural and chemical characterization**. High-resolution XRD measurements were performed using a four-circle Panalytical X'pert Pro diffractometer with Cu−K$\alpha$ radiation. XAS measurements were performed in a fluorescent mode at the ISS 8-ID beamline of National Synchrotron Light Source II at Brookhaven National Laboratory. The film microstructure was investigated with a Nion UltraSTEM 200 equipped with a fifth-order aberration corrector and a cold-field emission gun operated at 200 kV. The beam convergence half-angle was 30 mrad and the inner detector half-angle was 65 mrad. Electron energy-loss spectra were obtained with a collection half-angle of 48 mrad. The film strain distribution was analyzed by the GPA method.

**Electrical transport and magnetic property measurements.** Electrical transport measurements were performed using a Quantum Design physical property measurement system. In-plane transport on the films grown on STO substrates was measured using a typical Van der Pauw geometry. Out-of-plane transport for the films grown on Nb-STO substrates was measured using a two-terminal geometry with Nb-STO as the bottom electrode and a gold contact as the top electrode. Magnetic property measurements, including magnetic hysteresis loops and temperature-dependent magnetization, were performed using a 7 Tesla Quantum Design superconducting quantum interference device magnetometer. A 1000 Oe field in the in-plane direction was used for measuring M-T curves. The substrate background was removed for all measured samples.

**STM/STS measurements.** The STM/STS experiments were performed using an Omicron variable temperature AFM/STM system controlled by a SPECS Nanonis framework. All STS experiments were performed at 77 K with a base pressure less than $2\times10^{-10}$ Torr. Pt-Ir tips were calibrated on a clean Au(111) surface before measurements to ensure tip quality. Topographic images were collected in a constant current mode with a current setpoint of ~15 pA and tip bias of -2V (sample grounded). All the dI/dV spectra and line spectra were obtained using the lock-in technique with a modulation of 50 mV at 973 Hz on bias voltage. The thin film samples used for STM/STS measurements were transferred to a UHV-STM chamber immediately after growth. To remove any possible surface contamination, the samples were heated to 600 °C (20 °C/min) in 200 mTorr of $O_2$ for about 30 mins. *In situ* reflection high-energy electron diffraction (RHEED) was used to monitor the surface structure during the heating process until a clear crystalline RHEED pattern appeared, which indicates complete desorption of the amorphous surface contamination layer. The samples were quenched at 20 °C/min to room temperature and transferred in vacuum to a STM/STS analysis chamber.

**Acknowledgements**

This work was supported by the U.S. Department of Energy (DOE), Office of Science, Basic Energy Sciences (BES), Materials Sciences and Engineering Division. Part of this research was performed at the Center for Nanophase Materials Sciences at Oak Ridge National Laboratory, which is a DOE Office of Science User Facility. This research used resources of the Center for Functional Nanomaterials and the Inner Shell Spectroscopy 8-ID beamline of the National Synchrotron Light Source II, which are U.S. DOE Office of Science User Facilities at Brookhaven National Laboratory under Contract No. DE-SC0012704.

# Figures and Captions

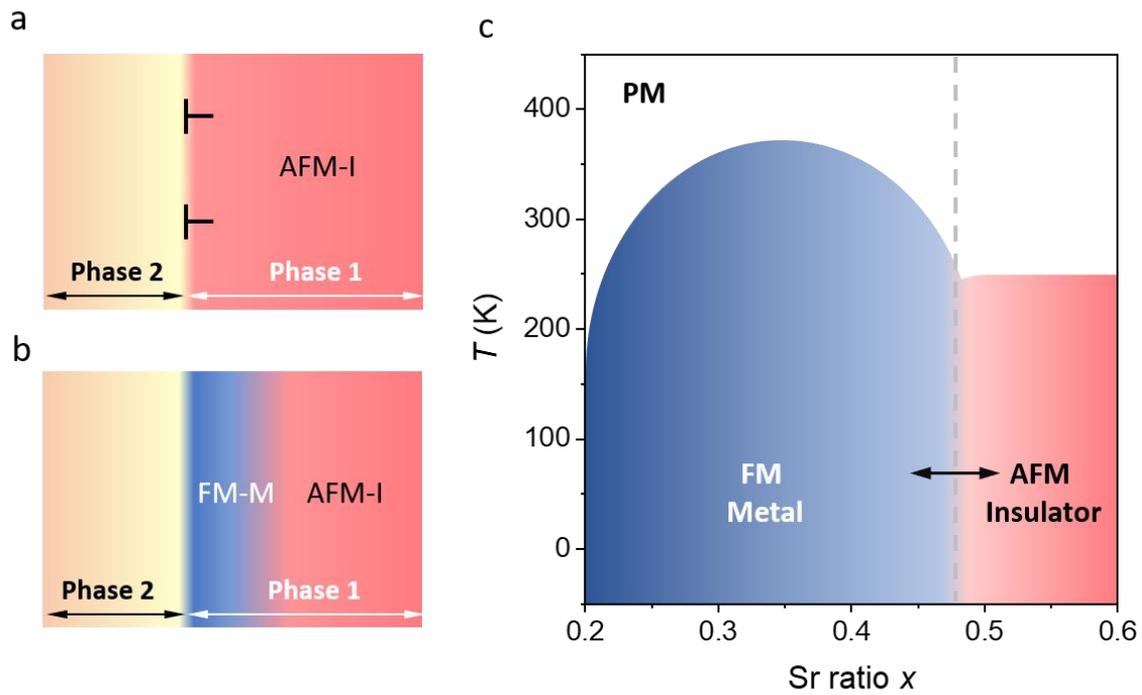

**Scheme 1**. Schematic diagram of (a) strain-relaxed and (b) chemically diffuse interface. (c) Compositional phase diagram of La$_{1-x}$Sr$_x$MnO$_3$ illustrating the metal-to-insulator phase transition boundary.

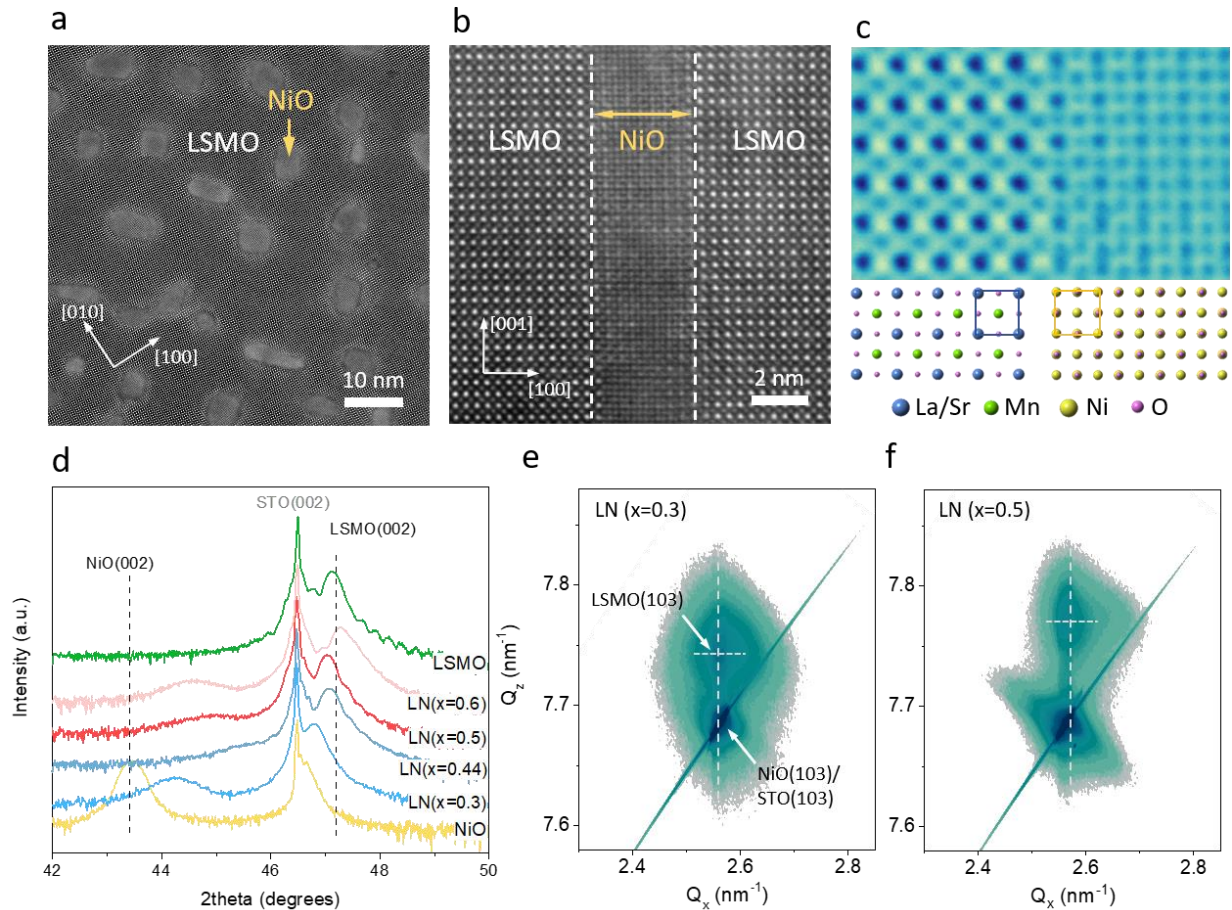

**Figure 1**. (a) Plan-view and (b) cross-sectional HAADF-STEM image of a LSMO-NiO ($x = 0.5$) VAN thin film demonstrating vertical interfaces formed by self-assembled NiO nanopillars in the planar LMSO matrix. (c) Atomic resolution HAADF-STEM image showing the epitaxial LSMO-NiO vertical interface with the corresponding crystallographic model. (d) $\theta-2\theta$ local XRD scans of LSMO-NiO (LN) VAN films with various Sr doping levels. Reciprocal space maps near the (103) STO diffraction peak for (e) LSMO-NiO ($x = 0.44$) and (f) LSMO-NiO ($x = 0.5$) VAN Films.

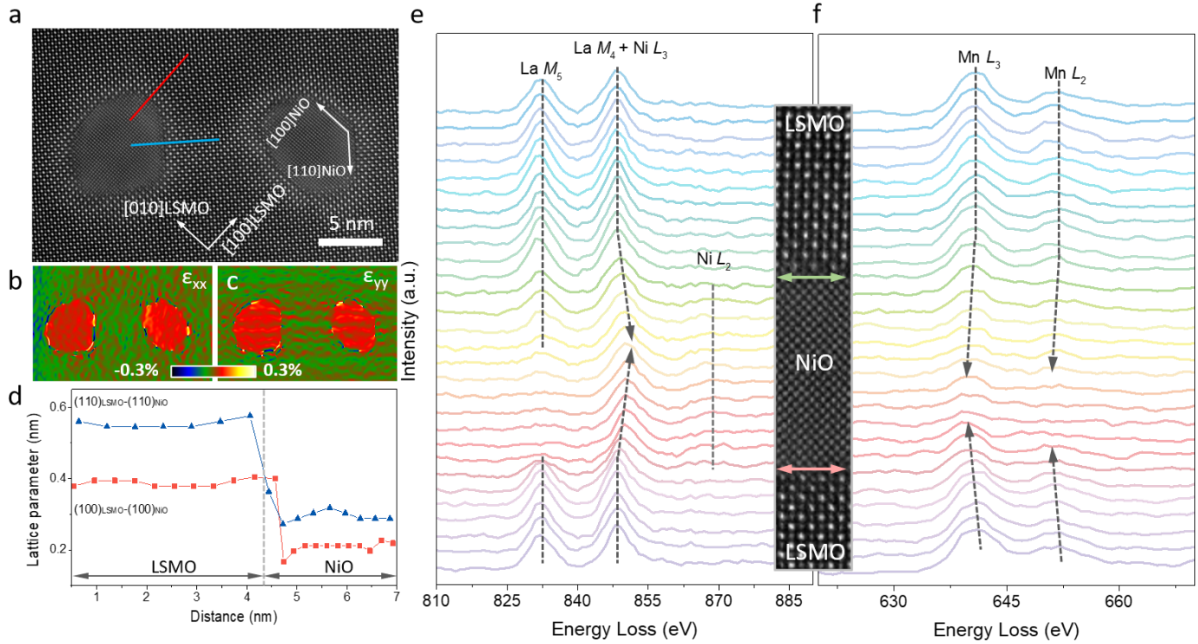

**Figure 2**. (a) Plan-view HAADF-STEM image of a LSMO-NiO ($x = 0.5$) VAN film showing in-plane lattice matching relations between LSMO and NiO. Corresponding strain maps in the (b) $\varepsilon_{xx}$ and (c) $\varepsilon_{yy}$ directions. (d) Lattice parameter variation across the interface region as marked by the red and blue lines in (a), which represent the epitaxial interface relationship of (100)LSMO-(100)NiO and (110)LSMO-(110)NiO, respectively. The error bar is 0.016 nm. Spatially-resolved EELS spectra of (e) La-*M* edge, Ni-*L* edge and (f) Mn-*L* edge across an entire NiO pillar, with the corresponding HAADF-STEM image turned sideways in the inset.

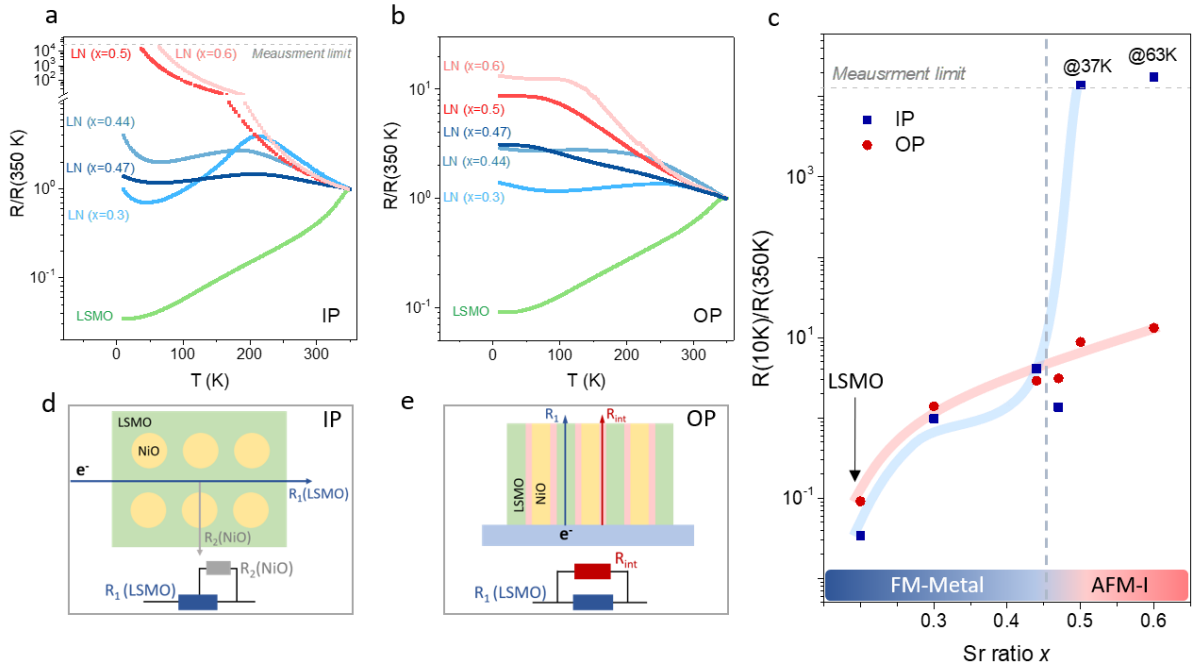

**Figure 3**. Temperature-dependent R/R(350K) of LSMO-NiO VAN films for (a) in-plane (IP) and (b) out-of-plane (OP) electrical transport. (c) R(10K)/R(350K) as a function of the Sr doping level for LSMO-NiO VAN films identifying remarkable out-of-plane conductivity for $x$ ranging between 0.47 and 0.6. Schematic and equivalent circuit for (d) in-plane and (e) out-of-plane electrical transport.

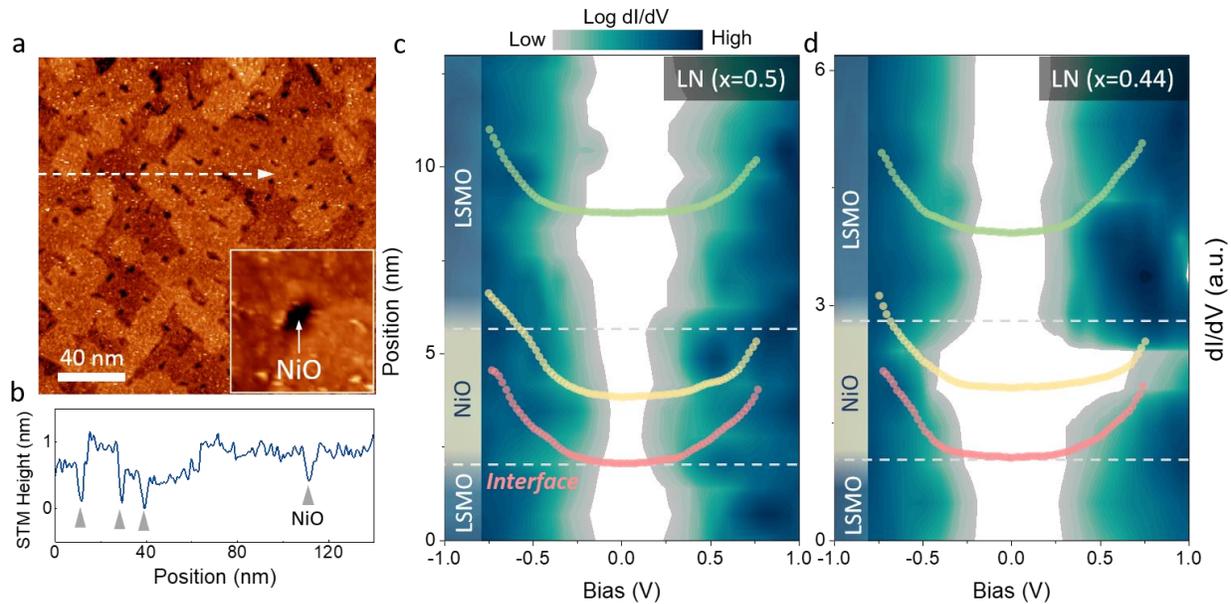

**Figure 4**. (a) STM image of a LSMO-NiO VAN film for $x = 0.5$ and (b) corresponding STM topography profile along the white line resolving clear NiO nanopillars on a smooth surface as marked by gray triangles. Contour map of spatially resolved dI/dV spectra from the surrounding LSMO matrix across a single NiO nanopillar for (c) LSMO-NiO ($x = 0.5$) and (d) LSMO-NiO ($x = 0.44$) VAN Films. The dashed lines mark the location of the interface region between LSMO and NiO. The individual differential conductivity spectra drawn on the contour maps are representative of the LSMO matrix (green), NiO nanopillars (yellow) and the vertical interface region (red).